# A hierarchy of thermal processes collapses under catalysis

Jeongrak Son[*] and Nelly H. Y. Ng[†]
*School of Physical and Mathematical Sciences, Nanyang Technological University, 637371, Singapore*

Thermal operations are a generic description for allowed state transitions under thermodynamic restrictions. However, the quest for simpler methods to encompass all these processes remains unfulfilled. We resolve this challenge through the catalytic use of thermal baths, which are assumed to be easily accessible. We select two sets of simplified operations: elementary thermal operations and Markovian thermal operations. They are known for their experimental feasibility, but fail to capture the full extent of thermal operations due to their innate Markovianity. We nevertheless demonstrate that this limitation can be overcome when the operations are enhanced by ambient-temperature Gibbs state catalysts. In essence, our result indicates that free states within thermal operations can act as catalysts that provide the necessary non-Markovianity for simpler operations. Furthermore, we prove that when any catalyst can be employed, different thermal processes (thermal operations, elementary thermal operations, and Markovian thermal operations) converge. Notably, our results extend to scenarios involving initial states with coherence in the energy eigenbasis, a notoriously difficult process to characterise.

Thermal operations (TO) [1] are a collection of quantum channels that capture the effects of heat exchange between a system and a thermal reservoir. These operations incorporate any evolution under the assumption that the system-reservoir composite is a closed system and that energy is strictly preserved. Due to its genericity, TO has been extensively studied in quantum thermodynamics [2–4] as a powerful tool for exploring how thermodynamics fundamentally deviates from the classical macroscopic picture, as one considers quantum systems [5–9]. These new discoveries impose fundamental limitations on the performance of quantum devices (such as efficiency or precision), and hence are critical for the optimization of device performance for the purposes of quantum information processing.

It remains a technical challenge to attain the fundamental bounds predicted by the TO framework in real-world implementations [10], as existing protocols [1, 5, 11, 12] for energetically-optimal TOs require demanding subroutines. Another hindrance arises from the structure of the theoretical development [13], which tends to focus on the possibility of state transformations, instead of providing concrete and simple heat baths and interaction Hamiltonians underlying the process. As a result, even when a transformation is known to be possible, it remains non-trivial to find the protocol that implements it [14, 15], obscuring the dynamical description of the process.

To address these challenges, a subset of TO, dubbed elementary thermal operations (ETO) [16], was proposed. It was originally envisioned that TO would be decomposed into sequences of much simpler ETO channels, involving at most two system levels at any given time. Such a goal would be analogous to having a small set of two-qubit universal gates in quantum computation. Since ETOs can be generally emulated through well understood interactions [17, 18], such a decomposition would provide a clearer implementation pathway to achieve the full extent of TO in practical, experimental settings. Moreover, by tracking states after each ETO, a time-resolved picture of the transformation emerges naturally, creating room for analysing dynamics [19]. Unfortunately, it was shown that not all TOs can be recast into a convex combination of ETOs, even when restricted to energy-incoherent transformations [16]. A seemingly much easier task of decomposing TO into series of TO involving at most all but one system levels at a time, is also shown to be impossible [20]. Since then, within the field, it has been generally assumed that it would be impossible to map of the complete set of thermal operations into simpler operations.

Two very recent results hinted at the potential of catalysis to narrow this gap between TOs and ETOs. Firstly, low-dimensional examples of recovering the full TO reachable set by adding a simple catalyst to ETOs were discovered [19]. Secondly, in the even more restricted subset of Markovian thermal operations (MTO) [21–23], Gibbs states which are assumed free in thermal operations, were found to be useful catalysts for transitions beyond MTO [24]. This counter-intuitive finding reveals a key insight that Gibbs states are useful as catalysts whenever the set of thermal processes have, to some extent, an in-built Markovian behaviour, i.e. some information is inevitably lost to the environment. While MTO is fully Markovian by design, ETO captures some non-Markovian behaviour because they employ finite-size baths and strong couplings. Yet, the extent of non-Markovianity is limited – after each ETO, bath states are discarded and a new Gibbs state retaining no memory is introduced for the subsequent step.

Is the inability of ETO to emulate TOs solely attributed to their Markovianity? If this were the case, injecting sufficient non-Markovianity should enable the implementation of any complex, multi-level TOs using only sequential two-level unitaries, which would be a significant simplification. In this work we answer this question affirmatively – indeed, any TO state transition can be accomplished with this strategy. Consequently, we discover that the use of Gibbs catalysts is the decisive factor in closing the gap between TOs and ETOs. More significantly, we show that Gibbs catalysts also close the gap between *catalytic* versions of TO and ETO. This represents significant progress in characterising catalytic ETOs, which has been an inaccessible problem due to the lack of efficiently computable state transition conditions in the ETO framework, except for special classes of initial states [19].

## PRELIMINARIES

### Elementary thermal operations

We briefly introduce the various classes of thermal processes, with different levels of non-Markovianity, and their relationship to one another. We denote the Hilbert spaces of system, bath and catalyst as $\mathcal{H}_S, \mathcal{H}_R$, and $\mathcal{H}_C$ respectively. The set of quantum states on Hilbert space $\mathcal{H}$ is denoted as $\mathcal{S}(\mathcal{H})$, and the set of linear operators as $\mathcal{L}(\mathcal{H})$. For any Hilbert space $\mathcal{H}_X$, we define an orthonormal energy eigenbasis $\{|i, g\rangle_X\}_{i,g}$ given by the Hamiltonian $H_X$, where the index $i$ denotes the energy eigenvalue $E_i$, and $g$ captures the degeneracy of that energy level. Whenever the spectrum of $X$ is non-degenerate, we drop the index $g$. We further denote the set of all reachable states $\{\phi | \rho \xrightarrow{X} \phi\}$ by a class of operations X as $\mathcal{R}_X(\rho)$. Acronyms for each operation can be found in Table I. For the main classes of thermal processes (TO, ETO, and MTO) relevant for this work, we have [16, 22, 23],

$$\mathcal{R}_{\mathrm{MTO}}(\rho), \mathcal{R}_{\mathrm{ETO}}(\rho) \subset \mathcal{R}_{\mathrm{TO}}(\rho), \quad (1)$$

and $\mathcal{R}_{\mathrm{MTO}}(\rho) \subset \mathcal{R}_{\mathrm{ETO}}(\rho)$ for energy-incoherent states $\rho$ [21].

Let us begin with the paradigm of elementary thermal operations [16], which can be decomposed into a series of two-level thermal operations.

**Definition 1** (ETO). An ETO sequence $\Phi : \mathcal{S}(\mathcal{H}_S) \to \mathcal{S}(\mathcal{H}_S)$ can be written as a series of channels, $\Phi = \Phi_N \circ \Phi_{N-1} \cdots \circ \Phi_1$, such that each $\Phi_i$ can be unravelled into three steps:

1. append a Gibbs state $\tau^\beta(H_{R_i}) = e^{-\beta H_{R_i}}/Z_{R_i} \in \mathcal{S}(\mathcal{H}_{R_i})$ with a suitable Hamiltonian $H_{R_i} \in \mathcal{L}(\mathcal{H}_{R_i})$ and corresponding partition function $Z_{R_i} = \mathrm{Tr}[e^{-\beta H_{R_i}}]$,

2. apply an energy-preserving unitary $V_{k_i l_i} \in \mathcal{L}(\mathcal{H}_S \otimes \mathcal{H}_{R_i})$ to the state $\rho \otimes \tau^\beta(H_{R_i})$, such that each $V_{k_i l_i}$ acts non-trivially on *at most* two system energy levels $|k_i\rangle_S, |l_i\rangle_S$. Note that $V_{k_i l_i}$ potentially involves many other bath energy levels. Nonetheless we refer to it as a *two-system-level unitary* due to its effect on $S$. Finally,

3. trace out the bath part using $\mathrm{Tr}_{R_i}$.

To summarise, each channel $\Phi_i$ is given by

$$\Phi_i(\varrho) = \mathrm{Tr}_{R_i}\left[V_{k_i l_i}\left(\varrho \otimes \tau^\beta(H_{R_i})\right)\left(V_{k_i l_i}\right)^\dagger\right]. \quad (2)$$

The definitions of TO and MTO are less directly relevant for the statement of our results, and therefore are left to the Supplemental Material. A generic TO has the same form as Eq. (2), with $V_{kl}$ replaced by a general energy-preserving unitary $U$. The critical difference is that for an ETO concatenation, each $\Phi_i$ in the sequence is implemented with a *fresh* bath $\tau^\beta(H_{R_i})$.

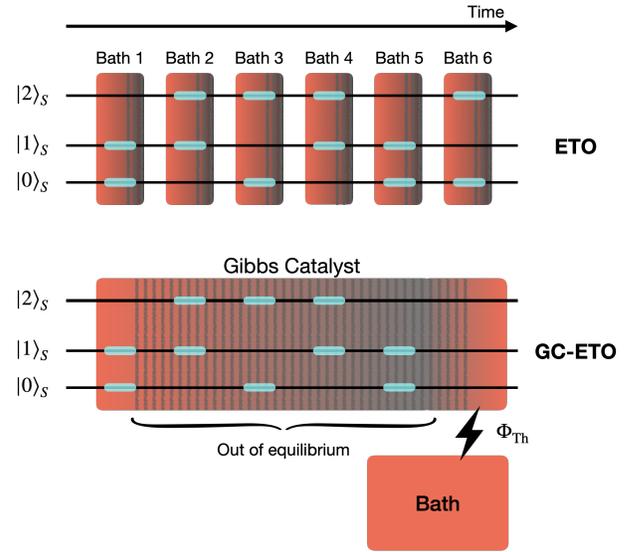

FIG. 1. Comparison between ETO and GC-ETO. The top diagram depicts an ETO sequence applied to a qutrit. At each step, two levels (highlighted in light blue) interact with a refreshed thermal bath (red), which becomes athermal at the end (grey). When two new levels are chosen, a fresh bath is also chosen. The lower part portrays a GC-ETO process with a catalyst starting from a Gibbs state (in red). During the process, the catalyst goes out of equilibrium (in grey with stripes), but at the end of the process, it is rethermalised via the thermalising channel $\Phi_{\mathrm{Th}} \in \mathrm{ETO}$.

Intuitively, the irreversible loss of information into bath $R_i$ in each channel $\Phi_i$ produces an in-built Markovian behaviour. Therefore, it is unsurprising that $\mathcal{R}_{\mathrm{ETO}}(\rho) \subsetneq \mathcal{R}_{\mathrm{TO}}(\rho)$ with a strict gap, for general inputs $\rho$ [25]. However, ETO is not fully Markovian either; each bath $\tau^\beta(H_{R_i})$ goes out of equilibrium during individual evolutions $\Phi_i$, in contrast to MTO.

### Elementary thermal operations with Gibbs state catalysts

To retrieve the information from the thermal reservoir, we must maintain full control over the baths used, rather than tracing them out after each two-system-level operation. We do so by modelling such baths as catalysts; we thus retain the baths throughout the process without tracing them out midway. Therefore, we inspect ETOs supplemented with Gibbs state catalysts, i.e. controllable thermal baths (see Fig. 1).

**Definition 2** (GC-ETO). A transformation $\rho \to \sigma$ is achievable by Gibbs-catalytic elementary thermal operations if there exists a Gibbs state $\tau^\beta(H_C) \in \mathcal{S}(\mathcal{H}_C)$, such that $\rho \otimes \tau^\beta(H_C) \xrightarrow{\mathrm{ETO}} \sigma \otimes \tau^\beta(H_C)$. In other words, $\sigma \otimes \tau^\beta(H_C) \in \mathcal{R}_{\mathrm{ETO}}\left(\rho \otimes \tau^\beta(H_C)\right)$.

However, we have not addressed one crucial difference between non-Markovianity and catalysis. Strict catalytic operations, as defined in Def. 2, require the catalyst to return to its original state, without any correlation with the system. Following a non-Markovian evolution, the system and the catalyst become correlated, and the catalyst state is generally no longer a Gibbs state. We resolve this discrepancy by noting that the full thermalisation can be performed by a sequence of ETOs (or MTOs), with the aid of an additional bath. With this, the restoration of the catalyst can always be done for free. Furthermore, since this channel transforms every density matrix into a single fixed point, it also eliminates any external correlations with the system.

| Free operations | With Gibbs catalysts | With any catalysts (Defs. 6, 7, and 8) |
|---|---|---|
| TO (Def. 9) | TO (Gibbs states are free) | CTO |
| ETO (Def. 1) | GC-ETO (Def. 2) = TO, Thm. 3 | CETO = CTO, Thm. 4 |
| MTO (Def. 10) | GC-MTO = TO, Thm. 3 | CMTO = CTO, Thm. 4 |

TABLE I. Various choices of free operations for thermodynamic resource theories, and their relationship with one another.

Gibbs states are free in TO and therefore do not provide any advantage to TO. In other words, GC-TO is equivalent to TO without any catalyst. Since, ETO is a subset of TO, GC-ETO is a subset of GC-TO, and in turn, a subset of TO. This means that

$$\mathcal{R}_{\text{GC-ETO}}(\rho) \subset \mathcal{R}_{\text{TO}}(\rho), \quad (3)$$

for any input state $\rho$. Similarly, we can establish the relation $\mathcal{R}_{\text{GC-MTO}}(\rho) \subset \mathcal{R}_{\text{TO}}(\rho)$ for any input state $\rho$.

## MAIN RESULTS

### Gibbs catalysts bridge ETOs and TOs

Our first result shows that the converse of Eq. (3) is also true.

**Theorem 3.** $\mathcal{R}_{\text{TO}}(\rho) = \mathcal{R}_{\text{GC-ETO}}(\rho) = \mathcal{R}_{\text{GC-MTO}}(\rho)$ *for all states $\rho$.*

*Proof.* We emphasise that this result holds for arbitrary *energy-coherent* initial states. For this, we only need to show that any TO channel can be recast into a GC-ETO (or GC-MTO) channel. Consider a TO channel $\Phi_{\text{TO}} : \mathcal{S}(\mathcal{H}_S) \to \mathcal{S}(\mathcal{H}_S)$ that admits the dilated form $\Phi_{\text{TO}}(\rho) = \text{Tr}_R \left[ U \left( \rho \otimes \tau^\beta(H_R) \right) U^\dagger \right]$, with some bath Hamiltonian $H_R \in \mathcal{L}(\mathcal{H}_R)$ and an energy-preserving unitary $U \in \mathcal{L}(\mathcal{H}_S \otimes \mathcal{H}_R)$, i.e. $[U, H_0] = 0$, where we denote the non-interacting Hamiltonian $H_0 = H_S + H_R$.

The language of Lie group and Lie algebra is useful for our proof, and we briefly summarise them in the Methods section. The set of all energy-preserving unitaries $G$ given the Hamiltonian $H_0$ [Eq. (15) in Methods] forms a Lie group, where the corresponding Lie algebra $\mathfrak{g}$ is the set of energy preserving Hamiltonians (times $-i$). Let us choose a set of linearly independent anti-Hermitian operators $\{K_1, K_2, \cdots, K_L\}$ that generates the Lie algebra $\mathfrak{g}$. Then from Lemma 5 in Methods, any energy-preserving unitary $U$ is a product of finite number of exponentials of the form $\exp(Kt)$, where $K \in \{K_1, K_2, \cdots, K_L\}$ and $t \in \mathbb{R}$.

Now we show that each $K_i$ in the set can be chosen as an operator acting on at most two levels. Any Hermitian operator $H_{\text{int}} \in \mathcal{L}(\mathcal{H}_S \otimes \mathcal{H}_C)$ can be expanded in the energy eigenbasis as

$$H_{\text{int}} = \sum_{k,l,E,E',g,g'} H^{(klEE'gg')}, \quad (4)$$

with each term $H^{(klEE'gg')} \propto |k\rangle\langle l|_S \otimes |E,g\rangle\langle E',g'|_C$ and $H^{(klEE'gg')} = (H^{(lkE'Eg'g)})^\dagger$. Here, we have chosen the catalyst to be the Gibbs state $\tau^\beta(H_C)$ with Hamiltonian $H_C = H_R$. Now observe that

$$[H^{(klEE'gg')}, H_0] = (E_k + E - E_l - E')H^{(klEE'gg')}, \quad (5)$$

where $E_i$ is the energy of level $|i\rangle_S$ and $E^{(')}$ denotes that of $|E^{(')}, g^{(')}\rangle_C$. Eq. (5) vanishes only when $E_k + E = E_l + E'$, i.e. when $|k\rangle_S |E,g\rangle_C$ and $|l\rangle_S |E',g'\rangle_C$ are in the same energy subspace. The energy-preserving condition $[H_{\text{int}}, H_0] = 0$ indicates that Eq. (5) is zero for all terms in Eq. (4). As a result, any energy-preserving Hamiltonian $H_{\text{int}}$ is a linear combination of (genuine) two-system-level terms

$$|k\rangle\langle l|_S \otimes |E - E_k, g\rangle\langle E - E_l, g'|_C + \text{h.c.}, \quad (6)$$
$$i|k\rangle\langle l|_S \otimes |E - E_k, g\rangle\langle E - E_l, g'|_C + \text{h.c.}, \quad (7)$$

with $k \neq l$ and one-system-level terms with $g \neq g'$,

$$|j\rangle\langle j|_S \otimes |E - E_j, g\rangle\langle E - E_j, g'|_C + \text{h.c.}, \quad (8)$$
$$i|j\rangle\langle j|_S \otimes |E - E_j, g\rangle\langle E - E_j, g'|_C + \text{h.c.}, \quad (9)$$
$$|j\rangle\langle j|_S \otimes |E - E_j, g\rangle\langle E - E_j, g|_C. \quad (10)$$

By collecting each of these terms (times $-i$), we obtain the basis set $\{K_1, K_2, \cdots, K_L\}$ of the Lie algebra $\mathfrak{g}$. Basis set is also a generating set; therefore, any energy-preserving unitary can be written as a finite product of at-most-two-level unitaries.

Since we regard the Gibbs state as a catalyst in GC-ETO framework, a unitary channel with the unitary operator $\exp(K_i t)$ is itself an ETO channel; the product of these unitaries is then an ETO sequence. After this sequence the resulting intermediate state is $\sigma_{SC}$, whose reduced state on $S$ is already the target state $\sigma$. Another ETO sequence that implements the fully thermalising channel yields the final product state $\sigma \otimes \tau^\beta(H_C)$. This concludes the proof of $\mathcal{R}_{\text{TO}}(\rho) \subset \mathcal{R}_{\text{GC-ETO}}(\rho)$.

We remark that the rank-1 projectors of the form Eq. (10), which act on a single system-catalyst level instead of two, generate valid ETO channels, according to the Definition 1. In addition, it is always possible to choose a different generating set $\{K'_1, K'_2, \cdots, K'_L\}$, where all elements are rank-2 operators; see the last section of Supplemental Material.

Finally, note that the same energy-preserving unitary channels generated by $K_i$ are also MTO channels since the system-catalyst composite is regarded as a controllable system. Furthermore, fully thermalising channel is also in MTO. Therefore, we also conclude that $\mathcal{R}_{\text{TO}}(\rho) \subset \mathcal{R}_{\text{GC-MTO}}(\rho)$. □

The above Theorem 3 states that for any TO channel, there always exists a GC-ETO process that emulates it. However, the explicit construction is not granted by the proof. We first provide an (approximate) generic decomposition of TO into Gibbs-catalytic operations, whose allowed interactions (times $-i$) constitute a basis $\{K_1, \cdots, K_L\}$ for the Lie algebra $\mathfrak{g}$ of the energy-preserving unitary group. This encompasses various scenarios, including but not limited to the cases of GC-ETO and GC-MTO. For even more general case, where $\{K_1, \cdots, K_L\}$ generates but does not give basis for $\mathfrak{g}$, see the last section of Supplemental Material. Trotter-Suzuki formula [26, 27] for generally non-commuting square matrices $A_i$ reads $e^{s \sum_i A_i} = (\prod_i e^{\frac{s}{M}A_i})^M + O(s^2/M)$, with the big-O notation $O(\cdot)$. The interaction Hamiltonian $H_{\text{int}} = \sum_j i r_j K_j \in i\mathfrak{g}$ with some real coefficients $r_j$. Applying the Trotter-Suzuki formula,

$$e^{-itH_{\text{int}}} = \left[\prod_j e^{\frac{t}{M}r_j K_j}\right]^M + O\left(\frac{t^2}{M}\right). \quad (11)$$

As $M \to \infty$, we have a construction of the GC-ETO (or other Gibbs-catalytic restricted operations) sequence approximating a given TO channel with vanishing error.

Interestingly, for the decomposition of TOs into GC-ETOs, an alternative exact construction exists. In Chapter 4.5.1 of Ref. [28], a method for decomposing a generic $d$-level unitary into $\frac{d(d-1)}{2}$ different 2-level unitaries is given. We observe that the argument also translates to *energy-preserving* unitaries. This is true, because any energy-preserving unitary can be first written as a direct sum of unitaries $\bigoplus_i U_i$ with each $U_i$ acting on a fully degenerate energy subspace. Using the construction in Ref. [28], we can decompose each $U_i$ into a series of two-level unitaries. Each of these two-level unitaries would be energy-preserving, and therefore the full decomposition corresponds to an ETO sequence.

### Example: cooling a qutrit

As a simple example that manifests Theorem 3, consider a qutrit state $\rho \in \mathcal{S}(\mathcal{H}_S)$ diagonal in the energy eigenbasis $\{|0\rangle_S, |1\rangle_S, |2\rangle_S\}$. The system Hamiltonian is set to be $H_S = E(|1\rangle\langle 1|_S + |2\rangle\langle 2|_S)$, satisfying $e^{-\beta E} = 1/2$. The population of $\rho$ is given as $\mathbf{p} = (0, 0.5, 0.5)$. The ground state has energy populations $\mathbf{q} = (1, 0, 0)$. It is known that a TO transition from $\mathbf{p}$ to $\mathbf{q}$ is feasible up to arbitrary precision, where precision increases with bath size [8]. However, via ETO, the maximum achievable ground state population is 0.75 [29].

We construct a GC-ETO process with a Gibbs state catalyst $\tau^\beta(H_C)$ that approaches the TO limit $\mathbf{q}$ with arbitrarily good accuracy. The catalyst Hamiltonian is chosen to be $H_C = \sum_{n=0}^{D-1} \sum_{j=1}^{\delta_n} nE|n, j\rangle\langle n, j|_C$, whose energy spectrum is equidistant up to the largest value $(D-1)E$. We choose the catalyst degeneracy to be $\delta_n = 2^n$ that grows exponentially with the energy. Furthermore, each degenerate energy subspace of the catalyst can always be divided into two subspaces of equal dimension for $n \geq 1$, i.e. $\delta_n/2 = 2^{n-1}$ is always an integer.

Now we define the elements of the interaction Hamiltonian for $SC$ such that half the terms are the interactions between $|0\rangle_S$ and $|1\rangle_S$, while the other half is between $|0\rangle_S$ and $|2\rangle_S$. More concretely,

$$H_{01}^{(n,k)} = |0\rangle\langle 1|_S \otimes |n, k\rangle\langle n-1, k|_C + \text{h.c.}, \quad (12)$$

$$H_{02}^{(n,k)} = |0\rangle\langle 2|_S \otimes |n, 2^{n-1} + k\rangle\langle n-1, k|_C + \text{h.c.}, \quad (13)$$

for $k = 1, \cdots, 2^{n-1}$. Each term is Hermitian, energy-preserving, and can generate unitaries $X_x^{(n,k)}$ that swap $|0\rangle_S |n, k + (x-1)2^{n-1}\rangle_C$ and $|x\rangle_S |n-1, k\rangle_C$ for $x = 1, 2$, while keeping all other states unchanged. Hence, we have now designed two-level unitaries $X_x^{(n,k)}$ on $SC$, and they commute with each other $[X_x^{(n,k)}, X_{x'}^{(n',k')}] = 0$, for all $x, x', n, n', k, k'$.

Starting from the initial state $\rho \otimes \tau^\beta(H_C)$, we may obtain the final, energy-incoherent state $\sigma = \text{Tr}_C[U(\rho \otimes \tau^\beta)U^\dagger]$, where $U = \prod_{n=1}^{D-1} \prod_{k=1}^{2^{n-1}} \prod_{x=0,1} X_x^{(n,k)}$. We label the energy population vector of $\sigma$ as $\mathbf{q}'$. Except for the populations of $I = \{|1\rangle_S |D-1, j\rangle_C, |2\rangle_S |D-1, j\rangle_C\}_{j=1}^{2^{D-1}}$, which remain untouched by $U$, all populations from $|1\rangle_S$ and $|2\rangle_S$ are now transferred to $|0\rangle_S$. Each invariant level in $I$ possesses the population $D^{-1}2^{-D}$. Therefore, the final system state population $\mathbf{q}' = (1 - D^{-1}, D^{-1}/2, D^{-1}/2)$ is obtained, and it approaches the TO limit as $D \to \infty$. Indeed, $\mathbf{q}'$ is the state with the highest ground state population achievable via TO using the bath with the Hamiltonian $H_C$.

In this construction, no Trotter error is incurred, since the two-level interaction terms commute. In fact, many pairs of two-system-level interaction terms would typically commute, since they act on different energy-subspaces. Using this property, a more efficient way of decomposing TO unitaries than Eq. (11) might be accessible.

### Equivalence between catalytic versions of thermal processes

Building on Theorem 3, we formulate our second main result: when arbitrary catalysts are allowed, the hierarchy of thermal processes – MTO, ETO, and TO, which were previously studied independently – collapses. Such a statement can be subtle due to the various types of catalysis [53], leading to significantly different state transition conditions. Nevertheless, since our result provides a direct and exact decomposition of the unitary corresponding to TO, it holds for any catalytic type (e.g. strict, approximate, or correlated), as long as the definition remains consistent across different thermal processes.

**Theorem 4.** $\mathcal{R}_{\text{CTO}}(\rho) = \mathcal{R}_{\text{CETO}}(\rho) = \mathcal{R}_{\text{CMTO}}(\rho)$ *for all $\rho$, where C may stand for either strict, approximate or correlated catalysis.*

*Proof.* From Eq. (1), we deduce that $\mathcal{R}_{\text{CETO}}(\rho), \mathcal{R}_{\text{CMTO}}(\rho) \subset \mathcal{R}_{\text{CTO}}(\rho)$. For the other direction, consider $\sigma \in \mathcal{R}_{\text{CTO}}(\rho)$: this implies the existence of a catalyst state $\mu_C$, such that $\sigma_{SC} \in \mathcal{R}_{\text{TO}}(\rho \otimes \mu_C)$ and $\text{Tr}_C[\sigma_{SC}] = \sigma$. Furthermore, the catalyst recovery condition, which depends on the definition of catalysis, is imposed – the reader is referred to the Methods section, in particular Definitions 6-8, where different catalytic types and their degrees of stringency for the catalyst recovery are elaborated. Using Theorem 3, we find that $\sigma_{SC} \in \mathcal{R}_{\text{GC-ETO}}(\rho \otimes \mu_C)$, or equivalently, $\sigma_{SCC'} := \sigma_{SC} \otimes \tau^\beta(H_{C'}) \in \mathcal{R}_{\text{ETO}}(\rho \otimes \mu_C \otimes \tau^\beta(H_{C'}))$ for some $H_{C'}$. Any catalyst recovery condition imposed on $\sigma_{SC}$ is satisfied by $\sigma_{SCC'}$; see Methods for more details. Therefore, we have that $\sigma \in \mathcal{R}_{\text{CETO}}(\rho)$, and hence $\mathcal{R}_{\text{CTO}}(\rho) \subset \mathcal{R}_{\text{CETO}}(\rho)$. The same proof strategy applies to MTO. □

### DISCUSSIONS AND CONCLUSIONS

We remark on the seemingly odd possibility of using Gibbs states, which are free for TO, as catalysts. Typically, in resource theories characterised by free operations $X$, $\tau$ being a free state indicates that the process

$$\rho \xrightarrow{X} \rho \otimes \tau \tag{14}$$

is always feasible for any $\rho$. Free states defined as such can never be a useful catalyst: if $\rho \otimes \tau \xrightarrow{X} \sigma \otimes \tau$, and if partial trace is also an allowed free operation, their concatenation $\rho \xrightarrow{X} \rho \otimes \tau \xrightarrow{X} \sigma \otimes \tau \xrightarrow{X} \sigma$ results in $\rho \xrightarrow{X} \sigma$, without the need of catalysis. To grasp the utility of Gibbs states as catalysts, we must observe that, for ETO and MTO, Gibbs states are *not* free according to the abovementioned notion. Although ETO permits the use of arbitrary bath Hamiltonians, each bath expires after a single step and thus cannot be integrated into the system as in Eq. (14). This is why the power of Gibbs catalysts was not well-known or studied before, with possibly the exception of Ref. [24].

In summary, we have proved that different thermal processes with a hierarchy – TO, ETO, and MTO – become equivalent when catalysts are employed. Specifically, all three of them allow the same set of state transitions under two conditions: *i)* when any ambient-temperature Gibbs states can be a catalyst, or *ii)* when any catalyst states and Hamiltonians are accessible. We emphasise two major implications of these results. Recall that the TOs specify the optimal thermal processes for given tasks; performance bounds, such as those for cooling [30, 31], work extraction [32–35], thermal engine operation [9, 36], or multi-copy state transformations [37–39], are known. Theorem 3 ensures that by embracing fractions of the given thermal environment as part of the controllable system, these limits can be attained by quantum devices operating only with simple and experiment-friendly building blocks, such as two-level operations or Markovian processes. Second, Theorem 4 fully characterises CETO and CMTO transitions when the input states are energy-incoherent, due to existing results in CTO transitions [40, 56], whereas even non-catalytic versions of ETO and MTO transitions do not have a computationally viable characterisation.

There are several interesting prospects for future studies. The proof of Theorem 3 showcases the versatility of our approach, which could inspire further investigations into novel elementary interactions capable of decomposing generic free operations. In particular, the proof would hold as long as we choose a set of unitaries that can achieve universality. This therefore establishes a fundamental and direct relationship between the ability of achieving (catalytic) thermal operations, to the ability of performing universal quantum computation. Moreover, Theorem 4 parallels the scenario when the locality is imposed as an additional restriction [41]. There, catalysis bypasses certain group-theoretic constraints [42] imposed upon the local symmetric operations and recover the fully global symmetric operations. In our work, the additional restriction was the number of levels a single step can manipulate, and the catalyst helps overcome this constraint by providing non-Markovianity. From these two complementary observations of catalytic power, we envisage more systematic studies on catalysis in relation to group-theoretic properties and non-Markovianity. In particular, we have bridged the size of the bath for state transitions to the size of the catalysts, for instance in our qutrit cooling example. This connection implies that results such as lower bounds on catalyst size [19, 43–46] and on bath size [6, 8, 14, 47–50] may directly translate into one another.

### METHODS

#### Lie groups and Lie algebras

We start by showing that the set of energy-preserving unitaries given the system and bath Hamiltonian, with dimensions $d_S$ and $d_R$, form a Lie group. Consider the set

$$G = \{U | U \in \text{U}(d_S d_R), [U, H_0] = 0\}, \tag{15}$$

consists of energy-preserving unitaries commuting with the total Hamiltonian $H_0 = H_S + H_R$. Then this group is a compact and connected subgroup of the unitary group $\text{U}(d_S d_R)$ and thus equivalent to the exponentials of anti-Hermitian matrices commuting with $H_0$ [15]. In other words,

$$G = \{e^K | K \in \mathcal{L}(\mathcal{H}_S \otimes \mathcal{H}_R), K^\dagger = -K, [K, H_0] = 0\}, \tag{16}$$

is a Lie subgroup of $\text{U}(d_S d_R)$. The corresponding Lie algebra is then immediately given by

$$\mathfrak{g} = \{-iH | H \in \mathcal{L}(\mathcal{H}_S \otimes \mathcal{H}_R), H^\dagger = H, [H_0, H] = 0\}, \tag{17}$$

the set of interaction Hamiltonians (times $-i$) that commute with $H_0$. Finally, we define the set $\mathcal{K} = \{K_1, K_2, \cdots, K_L\}$ that generates the Lie algebra $\mathfrak{g}$; that is, $\mathfrak{g}$ is the real linear span of a set consisting of all elements of $\mathcal{K}$ and their (repeated) commutators. The following lemma then holds for $G$ and $\mathfrak{g}$.

**Lemma 5** (Appendix D, Lemma 1 of [51]). *If a set $\{K_1, K_2, \cdots, K_L\}$ generates a Lie algebra $\mathfrak{g}$, any element $U$ of the corresponding connected Lie group $G$ can be expressed as*

$$U = \prod_{n=1}^{S} \exp(K_{i_n} t_n), \qquad (18)$$

*where $S \in \mathbb{N}$ is a finite number, $i_n \in \{1, \cdots, L\}$, and $t_n \in \mathbb{R}$.*

### Catalyst recovery conditions

An interesting way of extending the set of feasible state transitions with a fixed set of free operations is to introduce the concept of catalysis [52, 53]. Assume that one has access to an auxiliary state $\mu_C \in \mathcal{S}(\mathcal{H}_C)$. Instead of applying free operations only to the system of interest $\rho$, one can start from the larger state $\rho \otimes \mu_C$. If the catalyst $\mu_C$ can be retrieved after the operation, the process is catalytic. There are varying degrees of stringency in defining the 'recovery' of the catalyst [46, 54–59], and they affect the state transition conditions qualitatively. We will introduce three most prominent choices, following the nomenclature of Ref. [53].

In this section, we distinguish them by assigning different abbreviation: SC for strict catalysis, CC for correlated catalysis, and AC for approximate catalysis. In the main text, however, we assume that the definition of catalyst recovery condition is consistent for different thermal processes and denote all of them as CX for catalytic-X operations.

The most conservative category is the strict catalysis.

**Definition 6** (Strict catalysis). *Given a set of free operations $X$, we say that a transition $\rho \to \sigma$ via strict catalytic-$X$ is possible if there exist a catalyst state $\mu_C$ and a free operation $\Phi^{(X)} \in X$, such that*

$$\Phi^{(X)}(\rho \otimes \mu_C) = \sigma \otimes \mu_C. \qquad (19)$$

The set $\mathcal{R}_{\text{SCX}}(\rho)$ is a collection of all states $\sigma$ that are reachable via strict catalytic-$X$ from $\rho$, and it always includes $\mathcal{R}_X(\rho)$ as a subset.

In the proof of Theorem 4, if the imposed catalyst recovery condition follows Definition 6, we have the state after TO $\sigma_{SC} = \sigma \otimes \mu_C$, when $\sigma \in \mathcal{R}_{\text{SCTO}}(\rho)$ with a catalyst state $\mu_C$. Then, with a catalyst $\mu_C \otimes \tau^\beta(H_{C'})$, the ETO transition can produce a state $\sigma_{SCC'} = \sigma_{SC} \otimes \tau^\beta(H_{C'})$, i.e. $\sigma \otimes \mu_C \otimes \tau^\beta(H_{C'}) \in \mathcal{R}_{\text{ETO}}(\rho \otimes \mu_C \otimes \tau^\beta(H_{C'}))$. This satisfies the strict catlyst recovery condition in Definition 6.

Eq. (19) is a very strict condition; yet in many resource theories [19, 24, 40, 41, 60–64] it has been reported to be useful for enlarging the set of reachable states $\mathcal{R}_X$. For TO, inequalities including the entire family of Rényi divergences of the state with respect to the Gibbs state determine $\mathcal{R}_{\text{SCTO}}(\rho)$ for incoherent $\rho$ [40]. For ETO, however, due to the difficulty of characterising the reachable state set, only the simplest non-trivial cases of qutrit system and qubit catalyst have been studied [19]. In the main text, we prove that given the freedom of choosing any catalyst, the same state transition criteria for CTO can also be used for catalytic ETO, since $\mathcal{R}_{\text{SCTO}}(\rho) = \mathcal{R}_{\text{SCETO}}(\rho)$.

Having settled this, we turn to a second category for catalysis, known as correlated catalysis.

**Definition 7** (Correlated catalysis). *Given a set of free operations $X$, we say that a transition $\rho \to \sigma$ via correlated catalytic-$X$ is possible if there exist a catalyst state $\mu_C$ and a free operation $\Phi^{(X)} \in X$, such that $\Phi^{(X)}(\rho \otimes \mu_C) = \sigma_{SC}$, where*

$$\text{Tr}_C[\sigma_{SC}] = \sigma, \quad \text{Tr}_S[\sigma_{SC}] = \mu_C. \qquad (20)$$

The set $\mathcal{R}_{\text{CCX}}(\rho)$ is then a collection of all states $\sigma$ that are reachable via correlated catalytic-$X$ from $\rho$, and it always includes $\mathcal{R}_{\text{SCX}}(\rho)$ as a subset.

Similar to the case of strict catalysis, $\sigma \in \mathcal{R}_{\text{CCTO}}(\rho)$ implies that $\sigma_{SC} \in \mathcal{R}_{\text{TO}}(\rho \otimes \mu_C)$ with $\text{Tr}_C[\sigma_{SC}] = \sigma$ and $\text{Tr}_S[\sigma_{SC}] = \mu_C$. Appending a Gibbs state to prepare a larger catalyst $\mu_C \otimes \tau^\beta(H_{C'})$, the resulting state from an ETO transition $\sigma_{SCC'} = \sigma_{SC} \otimes \tau^\beta(H_{C'})$ satisfies the correlated catalyst recovery condition, since $\text{Tr}_{CC'}[\sigma_{SCC'}] = \sigma$ and $\text{Tr}_S[\sigma_{SCC'}] = \mu_C \otimes \tau^\beta(H_{C'})$. Eq. (20) allows for the retention of correlation between the system and the catalyst at the end of the process, as long as the catalyst reduced state is recovered. For thermal operations, it has been shown that the family of inequalities characterising SCTO transformations reduces to a single inequality involving the non-equilibrium free energy [56]. Again, $\mathcal{R}_{\text{CCTO}}(\rho) = \mathcal{R}_{\text{CCETO}}(\rho)$, and we have a complete characterisation for correlated catalytic ETOs.

Lastly, we turn to a much more relaxed version of catalysis, called approximate catalysis, where errors are allowed upon returning the catalyst.

**Definition 8** (Approximate catalysis). *Given a set of free operations $X$, we say that a transition $\rho \to \sigma$ via approximate catalytic-$X$ is possible if there exists a catalyst state $\mu_C$ and a free operation $\Phi^{(X)} \in X$, such that $\Phi^{(X)}(\rho \otimes \mu_C) = \sigma \otimes \mu'_C$, where*

$$d(\mu_C, \mu'_C) \leq \epsilon, \qquad (21)$$

for some distance measure $d$ and some prescribed parameter $\epsilon > 0$. Let us assume that the distance measure is subadditive with respect to tensor product, i.e. $d(\rho_1 \otimes \rho_2, \sigma_1 \otimes \sigma_2) \leq d(\rho_1, \sigma_1) + d(\rho_2, \sigma_2)$ for all $\rho_1, \rho_2, \sigma_1, \sigma_2$. The set $\mathcal{R}_{\text{ACX}}(\rho)$ is then a collection of all states $\sigma$ that are reachable via approximate catalytic-$X$ from $\rho$, and it always includes $\mathcal{R}_{\text{SCX}}(\rho)$ as a subset.

Suppose that $\sigma \in \mathcal{R}_{\text{ACTO}}(\rho)$ with a catalyst $\mu_C$, i.e. $\sigma \otimes \mu'_C \in \mathcal{R}_{\text{TO}}(\rho \otimes \mu_C)$. Then $\sigma \otimes \mu'_C \otimes \tau^\beta(H_{C'}) \in \mathcal{R}_{\text{ETO}}(\rho \otimes \mu_C \otimes \tau^\beta(H_{C'}))$ for some $H_{C'}$. From Eq. (21), the recovered catalyst satisfies $d(\mu'_C, \mu_C) \leq \epsilon$. Then, we also have $d(\mu'_C \otimes \tau^\beta(H_{C'}), \mu_C \otimes \tau^\beta(H_{C'})) \leq \epsilon$ from the subadditivity of the distance measure. In consequence, $\mathcal{R}_{\text{ACTO}}(\rho) = \mathcal{R}_{\text{ACETO}}(\rho)$.

### DATA AVAILABILITY

This work is purely analytic and no data were generated during the current study.

### ACKNOWLEDGEMENTS

This work was supported by the start-up grant of the Nanyang Assistant Professorship of Nanyang Technological University, Singapore. During the process of preparing this manuscript we were made aware of the overlap with [65], and we thank Jakub Czartowski, Alexssandre de Oliveira Junior, and Kamil Korzekwa for insightful discussions on this topic. We also thank Seok Hyung Lie and Frederik vom Ende for independent and constructive discussions and Marek Gluza for the careful reading of the manuscript.

### AUTHOR CONTRIBUTIONS

Both authors worked on developing the results and wrote the manuscript.

### COMPETING INTERESTS

The authors declare no competing interest.

---

* jeongrak.son@e.ntu.edu.sg
† nelly.ng@ntu.edu.sg

[1] D. Janzing, P. Wocjan, R. Zeier, R. Geiss, and T. Beth, Thermodynamic cost of reliability and low temperatures: Tightening Landauer's principle and the second law, Int. J. Th. Phys. **39**, 2717 (2000).

[2] F. G. S. L. Brandão, M. Horodecki, J. Oppenheim, J. M. Renes, and R. W. Spekkens, Resource theory of quantum states out of thermal equilibrium, Phys. Rev. Lett. **111**, 250404 (2013).

[3] G. Gour, M. P. Müller, V. Narasimhachar, R. W. Spekkens, and N. Yunger Halpern, The resource theory of informational nonequilibrium in thermodynamics, Phys. Rep. **583**, 1 (2015).

[4] N. H. Y. Ng and M. P. Woods, Resource theory of quantum thermodynamics: Thermal operations and second laws, in *Thermodynamics in the Quantum Regime: Fundamental Aspects and New Directions*, edited by F. Binder, L. A. Correa, C. Gogolin, J. Anders, and G. Adesso (Springer International Publishing, Cham, 2018) pp. 625–650.

[5] M. Horodecki and J. Oppenheim, Fundamental limitations for quantum and nanoscale thermodynamics, Nat. Commun. **4**, 1 (2013).

[6] D. Reeb and M. M. Wolf, An improved landauer principle with finite-size corrections, New J. Phys. **16**, 103011 (2014).

[7] N. Yunger Halpern, P. Faist, J. Oppenheim, and A. Winter, Microcanonical and resource-theoretic derivations of the thermal state of a quantum system with noncommuting charges, Nat. Commun. **7**, 12051 (2016).

[8] J. Scharlau and M. P. Mueller, Quantum Horn's lemma, finite heat baths, and the third law of thermodynamics, Quantum **2**, 54 (2018).

[9] M. P. Woods, N. H. Y. Ng, and S. Wehner, The maximum efficiency of nano heat engines depends on more than temperature, Quantum **3**, 177 (2019).

[10] N. Yunger Halpern, Toward physical realizations of thermodynamic resource theories, in *Information and Interaction: Eddington, Wheeler, and the Limits of Knowledge*, edited by I. T. Durham and D. Rickles (Springer International Publishing, Cham, 2017) pp. 135–166.

[11] C. Perry, P. Ćwikliński, J. Anders, M. Horodecki, and J. Oppenheim, A sufficient set of experimentally implementable thermal operations for small systems, Phys. Rev. X **8**, 041049 (2018).

[12] N. Shiraishi, Two constructive proofs on d-majorization and thermo-majorization, J Phys. A: Math. Theor. **53**, 425301 (2020).

[13] M. Lostaglio, An introductory review of the resource theory approach to thermodynamics, Rep. Prog. Phys. **82**, 114001 (2019).

[14] X. Hu and F. Ding, Thermal operations involving a single-mode bosonic bath, Phys. Rev. A **99**, 012104 (2019).

[15] F. vom Ende, Which bath Hamiltonians matter for thermal operations?, J. Math. Phys. **63**, 112202 (2022).

[16] M. Lostaglio, Á. M. Alhambra, and C. Perry, Elementary Thermal Operations, Quantum **2**, 52 (2018).

[17] E. Jaynes and F. Cummings, Comparison of quantum and semiclassical radiation theories with application to the beam maser, Proc. IEEE **51**, 89 (1963).

[18] V. Bužek, Jaynes-Cummings model with intensity-dependent coupling interacting with holstein-primakoff su(1,1) coherent state, Phys. Rev. A **39**, 3196 (1989).

[19] J. Son and N. H. Y. Ng, Catalysis in action via elementary thermal operations (2022), arXiv:2209.15213.

[20] P. Mazurek and M. Horodecki, Decomposability and convex structure of thermal processes, New J. Phys. **20**, 053040 (2018).

[21] M. Lostaglio and K. Korzekwa, Continuous thermomajorization and a complete set of laws for Markovian thermal processes, Phys. Rev. A **106**, 012426 (2022).

[22] G. Spaventa, S. F. Huelga, and M. B. Plenio, Capacity of non-markovianity to boost the efficiency of molecular switches, Phys. Rev. A **105**, 012420 (2022).

[23] F. vom Ende, E. Malvetti, G. Dirr, and T. Schulte-Herbrüggen, Exploring the limits of controlled markovian quantum dynamics with thermal resources (2023), arXiv:2303.01891 [quant-ph].

[24] K. Korzekwa and M. Lostaglio, Optimizing thermalization, Phys. Rev. Lett. **129**, 040602 (2022).

[25] Conventionally, convex combinations of different ETO sequences are also allowed and will be used in the proof of Theorem 4. Yet, for Theorem 3 a sequence of ETO without convex combinations suffices.

[26] H. F. Trotter, On the product of semi-groups of operators, Proc. Am. Math. Soc. **10**, 545 (1959).

[27] M. Suzuki, Generalized Trotter's formula and systematic approximants of exponential operators and inner derivations with applications to many-body problems, Commun. Math. Phys. **51**, 183 (1976).

[28] M. A. Nielsen and I. L. Chuang, *Quantum Computation and Quantum Information: 10th Anniversary Edition* (Cambridge University Press, 2010).

[29] This is the case when levels $|0\rangle_S, |1\rangle_S$ and then levels $|0\rangle_S, |2\rangle_S$ interact maximally, for example. See [19] for a complete characterisation of $\mathcal{R}_{\text{ETO}}(\rho)$ for any three-dimensional incoherent state $\rho$.

[30] F. Clivaz, R. Silva, G. Haack, J. B. Brask, N. Brunner, and M. Huber, Unifying paradigms of quantum refrigeration: A universal and attainable bound on cooling, Phys. Rev. Lett. **123**, 10.1103/physrevlett.123.170605 (2019).

[31] F. Clivaz, R. Silva, G. Haack, J. B. Brask, N. Brunner, and M. Huber, Unifying paradigms of quantum refrigeration: Fundamental limits of cooling and associated work costs, Phys. Rev. E **100**, 10.1103/physreve.100.042130 (2019).

[32] O. C. O. Dahlsten, R. Renner, E. Rieper, and V. Vedral, Inadequacy of von neumann entropy for characterizing extractable work, New J. Phys. **13**, 053015 (2011).

[33] J. Åberg, Truly work-like work extraction via a single-shot analysis, Nat. Commun. **4**, 1925 (2013).

[34] D. Egloff, O. C. O. Dahlsten, R. Renner, and V. Vedral, A measure of majorization emerging from single-shot statistical mechanics, New Journal of Physics **17**, 073001 (2015).

[35] A. M. Alhambra, L. Masanes, J. Oppenheim, and C. Perry, Fluctuating work: From quantum thermodynamical identities to a second law equality, Phys. Rev. X **6**, 041017 (2016).

[36] H. Tajima and M. Hayashi, Finite-size effect on optimal efficiency of heat engines, Phys. Rev. E **96**, 012128 (2017).

[37] C. T. Chubb, M. Tomamichel, and K. Korzekwa, Beyond the thermodynamic limit: finite-size corrections to state interconversion rates, Quantum **2**, 108 (2018).

[38] K. Korzekwa, C. T. Chubb, and M. Tomamichel, Avoiding irreversibility: Engineering resonant conversions of quantum resources, Phys. Rev. Lett. **122**, 110403 (2019).

[39] C. T. Chubb, M. Tomamichel, and K. Korzekwa, Moderate deviation analysis of majorization-based resource interconversion, Phys. Rev. A **99**, 032332 (2019).

[40] F. Brandão, M. Horodecki, N. Ng, J. Oppenheim, and S. Wehner, The second laws of quantum thermodynamics, Proc. Natl. Acad. Sci. U.S.A. **112**, 3275 (2015).

[41] I. Marvian, Restrictions on realizable unitary operations imposed by symmetry and locality, Nat. Phys. **18**, 283 (2022).

[42] I. Marvian, (non-)universality in symmetric quantum circuits: Why abelian symmetries are special (2023), arXiv:2302.12466 [quant-ph].

[43] P. Lipka-Bartosik and P. Skrzypczyk, All states are universal catalysts in quantum thermodynamics, Phys. Rev. X **11**, 011061 (2021).

[44] P. Boes, N. H. Ng, and H. Wilming, Variance of relative surprisal as single-shot quantifier, PRX Quantum **3**, 010325 (2022).

[45] F. Ding, X. Hu, and H. Fan, Amplifying asymmetry with correlating catalysts, Phys. Rev. A **103**, 022403 (2021).

[46] R. Takagi and N. Shiraishi, Correlation in catalysts enables arbitrary manipulation of quantum coherence, Phys. Rev. Lett. **128**, 240501 (2022).

[47] A. E. Allahverdyan, K. V. Hovhannisyan, D. Janzing, and G. Mahler, Thermodynamic limits of dynamic cooling, Phys. Rev. E **84**, 041109 (2011).

[48] F. Ticozzi and L. Viola, Quantum resources for purification and cooling: fundamental limits and opportunities, Sci. Rep. **4**, 5192 (2014).

[49] L. Masanes and J. Oppenheim, A general derivation and quantification of the third law of thermodynamics, Nat. Commun. **8**, 14538 (2017).

[50] J. G. Richens, A. M. Alhambra, and L. Masanes, Finite-bath corrections to the second law of thermodynamics, Phys. Rev. E **97**, 062132 (2018).

[51] D. D'Alessandro, *Introduction to Quantum Control and Dynamics*, 2nd ed., Advances in Applied Mathematics (Chapman and Hall, New York, 2021).

[52] C. Datta, T. V. Kondra, M. Miller, and A. Streltsov, Catalysis of entanglement and other quantum resources (2022), arXiv:2207.05694 [quant-ph].

[53] P. Lipka-Bartosik, H. Wilming, and N. H. Y. Ng, Catalysis in quantum information theory (2023), arXiv:2306.00798 [quant-ph].

[54] M. P. Müller and M. Pastena, A generalization of majorization that characterizes shannon entropy, IEEE Trans. Inf. Theory **62**, 1711 (2016).

[55] H. Wilming, R. Gallego, and J. Eisert, Axiomatic characterization of the quantum relative entropy and free energy, Entropy **19**, 10.3390/e19060241 (2017).

[56] M. P. Müller, Correlating thermal machines and the second law at the nanoscale, Phys. Rev. X **8**, 041051 (2018).

[57] P. Boes, H. Wilming, R. Gallego, and J. Eisert, Catalytic quantum randomness, Phys. Rev. X **8**, 041016 (2018).

[58] T. V. Kondra, C. Datta, and A. Streltsov, Catalytic transformations of pure entangled states, Phys. Rev. Lett. **127**, 150503 (2021).

[59] S. H. Lie and N. H. Y. Ng, Catalysis always degrades external quantum correlations (2023), arXiv:2303.02376 [quant-ph].

[60] D. Jonathan and M. B. Plenio, Entanglement-assisted local manipulation of pure quantum states, Phys. Rev. Lett. **83**, 3566 (1999).

[61] S. Daftuar and M. Klimesh, Mathematical structure of entanglement catalysis, Phys. Rev. A **64**, 042314 (2001).

[62] P. H. Anspach, Two-qubit catalysis in a four-state pure bipartite system (2001), arXiv:quant-ph/0102067 [quant-ph].

[63] M. Klimesh, Inequalities that collectively completely characterize the catalytic majorization relation (2007), arXiv:0709.3680 [quant-ph].

[64] K. Bu, U. Singh, and J. Wu, Catalytic coherence transformations, Phys. Rev. A **93**, 042326 (2016).

[65] J. Czartowski, A. de Oliveira Junior, and K. Korzekwa, Thermal recall: Memory-assisted markovian thermal processes (2023), arXiv:2303.12840 [quant-ph].

[66] E. Chitambar and G. Gour, Quantum resource theories, Rev. Mod. Phys. **91**, 025001 (2019).

[67] E. Ruch, R. Schranner, and T. H. Seligman, The mixing distance, J. Chem. Phys. **69**, 386 (1978).

[68] F. vom Ende and G. Dirr, The d-majorization polytope, Linear Algebra Appl. **649**, 152 (2022).

[69] A. de Oliveira Junior, J. Czartowski, K. Życzkowski, and K. Korzekwa, Geometric structure of thermal cones, Phys. Rev. E **106**, 064109 (2022).

[70] M. Lostaglio, D. Jennings, and T. Rudolph, Description of quantum coherence in thermodynamic processes requires constraints beyond free energy, Nat. Commun. **6**, 6383 (2015).

[71] P. Ćwikliński, M. Studziński, M. Horodecki, and J. Oppenheim, Limitations on the evolution of quantum coherences: Towards fully quantum second laws of thermodynamics, Phys. Rev. Lett.

**115**, 210403 (2015).
[72] M. Lostaglio, K. Korzekwa, D. Jennings, and T. Rudolph, Quantum coherence, time-translation symmetry, and thermodynamics, Phys. Rev. X **5**, 021001 (2015).
[73] G. Gour, Role of quantum coherence in thermodynamics, PRX Quantum **3**, 040323 (2022).
[74] P. Faist, J. Oppenheim, and R. Renner, Gibbs-preserving maps outperform thermal operations in the quantum regime, New J. Phys. **17**, 043003 (2015).
[75] Y. Ding, F. Ding, and X. Hu, Exploring the gap between thermal operations and enhanced thermal operations, Phys. Rev. A **103**, 052214 (2021).
[76] I. Marvian and R. W. Spekkens, Extending Noether's theorem by quantifying the asymmetry of quantum states, Nat. Commun. **5**, 3821 (2014).

# Supplemental Material for
# "A hierarchy of thermal processes collapses under catalysis"

Jeongrak Son[*] and Nelly H. Y. Ng[†]

*School of Physical and Mathematical Sciences, Nanyang Technological University, 637371, Singapore*

## THERMODYNAMIC RESOURCE THEORIES

In this section, we provide a concise background to the various thermal processes. In particular, for the unfamiliar reader, we briefly introduce the resource-theoretic approach to quantum thermodynamics and two special classes of thermal processes: thermal operations (TO), which encompass a very generic heat exchange process, and Markovian thermal operations (MTO), representing a much smaller, physically motivated set of processes.

Resource theories [66] are mathematical frameworks that have introduced a systematic approach for studying restrictions on operations. These restrictions can result from fundamental limitations such as conservation laws, or practical considerations such as the ease of implementing certain processes for an experimenter. In the presence of such restrictions, mathematically interesting structures, e.g. pre-ordering relations, emerge to govern the possibility of state transitions. If one can prepare $\rho'$ starting from $\rho$ while adhering to the aforementioned restrictions, then we say that $\rho$ is more resourceful than $\rho'$.

The resource theory of thermodynamics [2–4, 13] assumes that the (infinitely large) environment is fully thermal with a fixed temperature of $1/\beta$, and one can perform strictly energy-preserving operations between the system and the bath. Then, the used baths can be rethermalised for free using the remaining portion of the environment. When no further constraints are placed on the bath, this defines the set of TO channels [1, 5].

**Definition 9** (TO). A channel $\Phi : \mathcal{S}(\mathcal{H}_S) \to \mathcal{S}(\mathcal{H}_S)$ is a thermal operation if it admits the dilation

$$\Phi(\rho) = \mathrm{Tr}_R \left[ U \left( \rho \otimes \tau^\beta(H_R) \right) U^\dagger \right], \qquad (22)$$

for some bath Hamiltonian $H_R$ and energy-preserving unitary $U \in \mathcal{L}(\mathcal{H}_S \otimes \mathcal{H}_R)$ such that $[U, H_S + H_R] = 0$. Here, $\tau^\beta(H_R) = e^{-\beta H_R}/Z_R$ is a Gibbs state with respect to Hamiltonian $H_R$ and inverse temperature $\beta$.

We can also define thermal operations more generally to map states in $S$ to $S'$, by tracing out some $R'$ instead of $R$. Yet, for simplicity, we follow the vast majority of literature and present the case where the mapping is an endomorphism. Despite there being a large degree of freedom for choosing the bath Hamiltonian and unitary operations, there does not exist a sole quantifier (e.g. the non-equilibrium free energy) that can capture all the resources in the framework. Instead, there is a more stringent set of conditions for state transformations. For initial state $\rho$ that is energy-incoherent, i.e. block-diagonal in its energy eigenbasis, the set of reachable states $\mathcal{R}_{\mathrm{TO}}(\rho)$ via TO is well characterised by the criterion of thermomajorization [5, 67–69], which is easily computable.

However, for general $\rho$ that might be coherent, finding the necessary and sufficient condition for state transitions under TO is a long-standing open problem [70–73].

Since TO limits neither the size of baths nor the controllability of the system-bath composite, many of the energy-preserving unitaries are extremely demanding to execute. Recently, variants of free operations obtained by further restricting certain features of TO have been proposed to mitigate the experimental challenges. Elementary thermal operations (ETO) [16] (Def. 1 in the main text) constrain the number of system levels we have to control at a time, to be at most two.

Another notable class is MTO [21–24], which is the set of TO channels implementable using a fully Markovian bath. More specifically, suppose that an MTO is performed by a (time-dependent) energy-preserving interaction between the system and bath during time $[0, t]$. At any time $0 \leq t_1 < t$, the Markovianity assumption implies that the bath remains in equilibrium – this is generally the case when an infinite bath is used or when the timescale for bath rethermalization is much shorter compared to the system-bath interaction timescale. Consequently, the interaction during any interval $[t_1, t_2] \subset [0, t]$ should also qualify as an MTO, since the interaction is energy-preserving and the initial state starts from a system-bath product state.

The rigorous definition of MTO is as follows [22]:

**Definition 10** (MTO). A channel $\Phi_{(0,t)} : \mathcal{S}(\mathcal{H}_S) \to \mathcal{S}(\mathcal{H}_S)$ is a Markovian thermal operation if

1. $\Phi_{(0,t)}$ is a TO with the form

$$\Phi_{(0,t)} = \mathcal{T} \exp\left[ \int_0^t L[s] ds \right], \qquad (23)$$

for some generator $L[s]$ and time-ordering $\mathcal{T}$ and

2. for any interval $[t_1, t_2] \subset [0, t]$, the channel

$$\Phi_{(t_1,t_2)} = \mathcal{T} \exp\left[ \int_{t_1}^{t_2} L[s] ds \right], \qquad (24)$$

is also a TO.

Similarly, one can define slightly different versions such as Markovian thermal processes and Markovian Gibbs-preserving maps. These variants consider channels $\Phi$ that are not necessarily contained in TO. For example, Markovian thermal processes correspond to Markovian maps whose only restriction is to 1) preserve the Gibbs state $\tau^\beta(H_S)$, and 2) be time-translation covariant. These are also referred to as enhanced thermal operations [71]. In general, both enhanced thermal operations and Gibbs-preserving maps are known to be strictly larger sets compared to TO [74, 75]. We refer the reader to Ref. [23] for a more detailed analysis on Markovian versions of these classes. In fact, in [21, 24] the set of free operations is defined so that the channels always have a trajectory inside the enhanced thermal operations, instead of TOs. Yet, enhanced thermal operations coincide with TOs for energy-incoherent initial states, which is the assumption of [21, 24]. Hence, their results can still be applied to MTO. One such notable result is that for any energy-incoherent initial state $\rho$, the reachable states $\mathcal{R}_{\text{MTO}}(\rho)$ is fully achievable by concatenations of two-system-level MTO, which in turn gives the relation $\mathcal{R}_{\text{MTO}}(\rho) \subset \mathcal{R}_{\text{ETO}}(\rho)$.

# GIBBS CATALYTIC ELEMENTARY THERMAL OPERATION WITHOUT RANK-1 PROJECTORS

In the main text, we constructed an ETO sequence that reproduces an arbitrary TO using elementary unitary channels, including those generated by rank-1 projectors. Our definition of ETOs, which aligns with previous studies, states that each ETO unitary acts non-trivially on at most two levels; hence, these rank-1 projectors generate valid ETO channels. Nevertheless, we would like to demonstrate that even with a stricter definition, one that only includes genuinely two-level unitaries and excludes rank-1 generators, it is still possible to decompose any TO channel into a GC-ETO sequence.

We first reiterate the basis of energy-preserving interaction Hamiltonians (times $-i$)

$$h_E^{(k,l,g,g')} = -i|k\rangle\langle l|_S \otimes |E-E_k,g\rangle\langle E-E_l,g'|_C - i|l\rangle\langle k|_S \otimes |E-E_l,g'\rangle\langle E-E_k,g|_C, \tag{25}$$

$$m_E^{(k,l,g,g')} = |k\rangle\langle l|_S \otimes |E-E_k,g\rangle\langle E-E_l,g'|_C - |l\rangle\langle k|_S \otimes |E-E_l,g'\rangle\langle E-E_k,g|_C, \tag{26}$$

$$p_E^{(j,g)} = -i|j\rangle\langle j|_S \otimes |E-E_j,g\rangle\langle E-E_j,g|_C, \tag{27}$$

where $k = l$ or $g = g'$ are also allowed. Let us define the set $\mathcal{K}$ to be a collection of all such operators, i.e. any element in the Lie algebra $\mathfrak{g}$ is a linear combination of $\mathcal{K}$ elements. However, another (potentially smaller) set $\mathcal{K}'$ can generate the Lie algebra $\mathfrak{g}$. Observe that

$$f_E^{(k,l,g,g')} := \frac{1}{2}\left[h_E^{(k,l,g,g')}, m_E^{(k,l,g,g')}\right] = i|k\rangle\langle k|_S \otimes |E-E_k,g\rangle\langle E-E_k,g|_C - i|l\rangle\langle l|_S \otimes |E-E_l,g'\rangle\langle E-E_l,g'|_C. \tag{28}$$

By defining another rank-2 energy-preserving operator

$$g_E^{(k,l,g,g')} := i|k\rangle\langle k|_S \otimes |E-E_k,g\rangle\langle E-E_k,g|_C + i|l\rangle\langle l|_S \otimes |E-E_l,g'\rangle\langle E-E_l,g'|_C, \tag{29}$$

the rank-1 operator $p_E^{(j,g)}$ can be written as a linear combination of rank-2 operators as

$$p_E^{(j,g)} = -\frac{1}{2}\left(f_E^{(j,k,g,g')} + g_E^{(j,k,g,g')}\right). \tag{30}$$

The decomposition in Eq. (30) always exists when two or more (system-plus-catalyst) energy levels share the same energy $E$. To eliminate the potential subtleties that may arise from the non-degenerate energy subspaces, we assume that an additional two-dimensional thermal catalyst with Hamiltonian $H_{C'} = 0$ is appended to the original Gibbs state catalyst $\tau^\beta(H_C)$. Now any energy level $|j\rangle_S |E - E_j, g\rangle$ has at least one another level $|j\rangle_S |E - E_j, g'\rangle$ with the same energy $E$. This does not affect other parts of the proof, as any thermal operation executable with a bath $\tau^\beta(H_C)$ can also be performed with a larger bath $\tau^\beta(H_C) \otimes \tau^\beta(H_{C'})$.

Then, we define the set $\mathcal{K}'$, which is a collection of $h_E^{(k,l,g,g')}$, $m_E^{(k,l,g,g')}$, and $g_E^{(k,l,g,g')}$ in Eqs. (25), (26), and (29) for all $E, k, l, g, g'$, which are all rank-2 energy-preserving operators.

The Lie algebra generated by $\mathcal{K}'$ is the set of anti-Hermitian energy-preserving operators

$$-iH_{\text{int}} = \sum_j a_j K'_j + \sum_{j,k} b_{j,k}[K'_j, K'_k] + \sum_{j,k,l} c_{j,k,l}[K'_j, [K'_k, K'_l]] + \cdots \tag{31}$$

with the real coefficients $a_j, b_{j,k}, c_{j,k,l}, \cdots$ and $K'_j$, elements of the set $\mathcal{K}'$. From Eqs. (28) and (30), we observe that $\mathfrak{g}$ is the Lie algebra generated by $\mathcal{K}'$.

We now need a new (approximate) construction that is more general than Trotter-Suzuki formula, which can only exponentiate linear combinations of operators. From a version of Baker-Campbell-Hausdorff formulae, for any (non-commuting) square

matrices $A$ and $B$, $e^{-tA}e^{-tB}e^{tA}e^{tB} = e^{t^2[A,B]} + O(t^3)$ [51]. Then, we can approach

$$e^{t[K'_j, K'_k]} = \left(e^{-\sqrt{\frac{t}{M}}K'_j} e^{-\sqrt{\frac{t}{M}}K'_k} e^{\sqrt{\frac{t}{M}}K'_j} e^{\sqrt{\frac{t}{M}}K'_k}\right)^M + O\left(\frac{t^{3/2}}{M^{1/2}}\right), \quad (32)$$

by setting $M \to \infty$. Combining Eq. (32) with the Trotter-Suzuki formula, the general energy-preserving unitary $e^{-iH_{\text{int}}}$ with the interaction Hamiltonian of the form Eq. (31) can be decomposed into a product of unitaries of the form $e^{tK'}$. First, using the Trotter-Suzuki formula,

$$e^{-iH_{\text{int}}t} = \left(\prod_j e^{\frac{a_j}{M}K'_j} \prod_{j,k} e^{\frac{b_{j,k}}{M}[K'_j, K'_k]} \cdots\right)^M. \quad (33)$$

Second, we replace all exponentials of (repeated) commutators by exponentials of the form $e^{tK'}$ using Eq. (32) multiple times. Then for large $M$, we achieve the desired (approximate) construction for a GC-ETO decomposition of an arbitrary TO, with arbitrarily small error. Note that this construction is again general; any restricted operations, whose interaction Hamiltonians (times $-i$) are selected from the set $\mathcal{K}'$ that generates the Lie algebra $\mathfrak{g}$, can decompose any TO with the aid of Gibbs state catalysts.